\documentclass[12pt,preprint]{aastex}

\def\gta{\ifmmode {\mathbin{\lower 3pt\hbox   
    {$\,\rlap{\raise 5pt\hbox{$\char'076$}}\mathchar"7218\,$}}}
    \else {${\mathbin{\lower 3pt\hbox
    {$\rlap{\raise 5pt\hbox{$\char'076$}}\mathchar"7218\,$}}}
    $}\fi}
\def\lta{\ifmmode {\,\mathbin{\lower 3pt\hbox   
    {$\,\rlap{\raise 5pt\hbox{$\char'074$}}\mathchar"7218\,$}}}
    \else {${\mathbin{\lower 3pt\hbox
    {$\rlap{\raise 5pt\hbox{$\char'074$}}\mathchar"7218\,$}}}
    $}\fi}

\shorttitle {Thermonuclear flame spreading on neutron stars}
\shortauthors {Bhattacharyya \& Strohmayer}

\begin{document}

\title{Spreading of thermonuclear flames on the neutron star in SAX
J1808.4--3658: an observational tool}

\author {Sudip Bhattacharyya\altaffilmark{1,2}, and Tod
E. Strohmayer\altaffilmark{2}}

\altaffiltext{1}{Department of Astronomy, University of Maryland at
College Park, College Park, MD 20742-2421}

\altaffiltext{2}{X-ray Astrophysics Lab,
Exploration of the Universe Division,
NASA's Goddard Space Flight Center,
Greenbelt, MD 20771; sudip@milkyway.gsfc.nasa.gov,
stroh@clarence.gsfc.nasa.gov}

\begin{abstract}

We analyse archival Rossi X-Ray Timing Explorer (RXTE) proportional
counter array (PCA) data of thermonuclear X-ray bursts from the 2002
outburst of the accreting millisecond pulsar SAX J1808.4--3658. We
present evidence of nonmonotonic variations of oscillation
frequency during burst rise, and correlations among the time
evolution of the oscillation frequency, amplitude, and the inferred
burning region area. We also discuss that the amplitude and burning
region area evolutions are consistent with thermonuclear flame
spreading on the neutron star surface.  Based on this discussion, we
infer that for the 2002 Oct. 15 thermonuclear burst, the ignition
likely occured in the mid-latitudes, the burning region took $\sim
0.2$ s to nearly encircle the equatorial region of the neutron star,
and after that the lower amplitude oscillation originated from the
remaining asymmetry of the burning front in the same hemisphere where
the burst ignited.  Our observational findings and theoretical
discussion indicate that studies of the evolution of burst oscillation
properties during burst rise can provide a powerful tool to understand
thermonuclear flame spreading on neutron star surfaces under extreme
physical conditions.
\end{abstract}


\section {Introduction} \label{sec: 1}

X-ray bursts are produced by thermonuclear burning of matter
accumulated on the surfaces of accreting neutron stars (Woosley, \&
Taam 1976; Lamb, \& Lamb 1978). During many bursts, millisecond period
brightness oscillations are generated by the combination of rapid
stellar rotation and an asymmetric brightness pattern on the neutron
star surface (Strohmayer, \& Bildsten 2003).  The period of these
oscillations is very close to the stellar spin period (Chakrabarty et
al. 2003; Strohmayer et al. 2003). Moreover, as this timing feature
originates from the surface of the neutron star, its detailed modeling
may be useful to constrain stellar structure parameters, and hence the
equation of state models of the dense matter in the neutron star core
(Bhattacharyya et al. 2005; Miller, \& Lamb 1998; Nath, Strohmayer, \&
Swank 2002; Muno, \"Ozel, \& Chakrabarty 2002).  

Modeling of burst
oscillations can also be useful to understand neutron star
atmospheres, surface fluid motions, and for mapping the magnetic field
structure on the stellar surface. For example, the evolution of
frequency of these oscillations during the burst rise phase, may
provide information on the spreading of thermonuclear flames under the
extreme physical conditions that exist on neutron stars (e.g.,
Bhattacharyya \& Strohmayer 2005). This is because bursts almost
certainly ignite at a particular point on the stellar surface (as
simultaneous ignition over the whole surface would require very fine
tuning), and then spread to burn all the surface fuel (Fryxell \&
Woosley 1982; Cumming \& Bildsten 2000; Spitkovsky, Levin, \&
Ushomirsky 2002; Bhattacharyya \& Strohmayer 2006a). This slow
(compared to the rotational speed) movement and spreading of the
burning region (along with other physical effects such as the
increased scale height due to burning) may give rise to complex
frequency evolution of the observed burst oscillations.  This
spreading would also cause the observed burst intensity to increase,
and the oscillation amplitude to decrease. Moreover, the increase in
emission area can be estimated by spectral analysis (Strohmayer, Zhang
\& Swank 1997; Bhattacharyya \& Strohmayer 2006a; 2006b).  Therefore,
simultaneous modeling of the evolution of burst intensity, oscillation
frequency and amplitude, and spectral properties can, in principle, be
a powerful tool to understand the propagation of burning fronts on
neutron star surfaces under conditions of extreme radiative pressure,
magnetic field and gravity.

Frequency evolution during burst rise oscillations has so far been
observed from two low mass X-ray binary (LMXB) sytems: SAX
J1808.4--3658 and 4U 1636--536 (Chakrabarty et al. 2003; Bhattacharyya
\& Strohmayer 2005).  The Rossi X-ray Timing Explorer (RXTE) observed
the 401 Hz X-ray pulsar SAX J1808.4--3658 in October and November of
2002, when it was in outburst. Four type I X-ray bursts were detected
during these observations (Chakrabarty et al. 2003), three of which
showed strong millisecond period brightness oscillations during burst
rise. A previous study found that as the burst intensity rises, the
oscillation frequency also increases by $\sim 5$~Hz and may overshoot
the stellar spin frequency (Chakrabarty et al. 2003).  Here, we
analyse these archival data in order to model the time evolution of
different burst properties and search for correlations among them.

In our study we find evidence of nonmonotonic variations in the oscillation
frequency during the rising phase of bursts. This is the first report
of such variations from any source. The frequency modulation is
correlated with the evolution of oscillation amplitude and burning
region area.  In \S~3, we discuss that 
the correlated amplitude and area evolution is consistent with 
thermonuclear flame propagation on the neutron star surface.

\section {Data Analysis and Results} \label{sec: 2}

We analyse the archival RXTE proportional counter array (PCA) data of
the 2002 outburst from SAX J1808.4--3658. Three thermonuclear bursts
with significant millisecond oscillations during the rising phase are
found in the ObsIds: 70080-01-01-000 (Oct 15), 70080-01-02-000 (Oct
18), and 70080-01-02-04 (Oct 19). First we explore the frequency
evolution of burst rise oscillations during these bursts using three
procedures.  We calculate dynamic power spectra (Strohmayer \&
Markwardt 1999) with time sampling short enough to resolve the burst
rise interval, but large enough to accumulate sufficient signal power
above the noise level. The dynamic spectra (panel {\it a} of Fig. 1,
and Fig. 2) provide indications of the frequency evolution
behavior. To confirm the indications in the dynamic spectra, we carry
out a phase timing analysis (Muno et al. 2000). We divide the burst
rise time interval into several bins of a fixed chosen length, and
then assuming a frequency evolution model, we calculate the average
phase $(\psi_k)$ in each bin $(k)$. The corresponding $\chi^2$ is
calculated using the formula $\chi^2 = \sum^M_{k=1}
(\psi_k-\bar\psi_k)^2/\sigma^2_{\psi_k}$ (Strohmayer \& Markwardt
2002), where $M$ is the number of bins.  For this study we used
extensive burst rise simulations to evaluate the uncertainty,
$\sigma_{\psi_k}$, as a function of the $Z^2$ power in each time bin.
We find the best fit parameter values for various frequency evolution
models by minimizing $\chi^2$, and we calculate the uncertainty in
each parameter by finding the change which produces the appropriate
increase in $\chi^2$ (Strohmayer \& Markwardt 2002; Press et
al. 1992).  Finally, we calculate the total $Z^2$ power (Strohmayer \&
Markwardt 2002) for the entire rise interval (ie. without binning)
using the best fit frequency evolution model parameters, and ensure
that this power is close to the maximum power obtained from any
parameter values.

We first fit the oscillations during the rising phase of the Oct. 15
burst with a constant frequency model. This gives a best fit frequency
of $400.91$~Hz and $\chi^2$/dof $= 30.04/8$. We evaluate the
significance of this $\chi^2$ value using simulations, and find a
probability of 0.005 to obtain such a value by chance. Since the
constant frequency model does not describe the data well, we next
consider more complex models. These are; (1) linear frequency
increase, (2) second order polynomial, and (3) linear increase and
subsequent linear decrease models (see Table 1 for a description of
the models). These models give $\chi^2$/dof values of $17.73/7$,
$17.35/6$, and $11.68/5$ respectively. Although these models give
better fits, they still have reduced $\chi^2$ values $> 2$, and do not
describe the data very well. A slightly more complex model which fits
the data well has a linear frequency increase followed by a second
order polynomial, and gives a $\chi^2$/dof value of 3.36/4. Of the
models tested we consider this the best description of the Oct. 15
burst, and the best fit parameter values are given in Table 1.  Panel
{\it a} of Fig. 1 shows that this model is consistent with the dynamic
power contours, and it indicates a frequency increase (by a few Hertz)
for the first $\sim 0.2$ s from burst onset, then a frequency decrease
(by $\sim 1$~Hz), and a subsequent increase.  From Table 1, we note
that the model parameters $\nu_{\rm 0}$ \& $\dot\nu_{\rm 1}$ are
practically unconstrained from the lower and upper sides respectively.
This implies that the initial frequency increase can be well fit by a
very steep model. However, the other sides of these parameters are
reasonably well constrained, which shows that a constant frequency
model is insufficient to model this portion.  Next we fit the rise
oscillations from the Oct. 18 \& Oct. 19 bursts with the same
frequency evolution models. The constant frequency model for the
Oct. 18 burst gives a $\chi^2$/dof $= 154.51/8$, and hence can be
strongly rejected. This confirms the conclusions of Chakrabarty et
al. (2003) who first noted the large frequency increase present during
this burst. A linear frequency increase model gives a $\chi^2$/dof $=
19.35/7$ that, though better, is still uncomfortably large to be
acceptable.  The next more complex model (a constant frequency, \&
subsequent linear increase) gives a $\chi^2$/dof $= 8.44/6$ (see Table
1, and the upper panel of Fig. 2), and is statistically
acceptable. However, we note that the dynamic power contours (upper
panel, Fig. 2) are suggestive of an initial frequency increase,
decrease, and increase behavior (shown by the dotted curve)
qualitatively similar to that for the Oct. 15 burst. Although higher
signal to noise ratios per time bin would likely be required to
confirm this behavior, this (dotted curve) model does give a higher
total $Z^2$ power than that given by the best fit $\chi^2$ model
(solid curve) in Table 1.  
Finally, the constant frequency model for the Oct. 19 burst gives a
$\chi^2$/dof $= 22.03/8$, which is also unacceptably high. A linear
frequency increase model (see Table 1, and the lower panel of Fig. 2)
gives a $\chi^2$/dof $= 6.46/7$, and is acceptable for this burst.
We note that although the oscillation frequency behavior of the
Oct. 19 burst is different from that of the Oct. 15 \& 18 bursts, the
former burst seems otherwise similar to the latter ones.  For example,
all three bursts show photospheric radius expansion, and their rise
times and durations are $\le 1$ s and a few tens of seconds
respectively. Thus, the variations in inferred frequency evolution
must be associated with some variable which does not drastically alter
the gross properties of the burst. One possible variable might be the
initial lattitude of ignition, although large variations in this
quantity might be expected to affect other burst properties, such as
the rise time, as well. Panel {\it b} of Fig. 1 shows the rms
amplitude variation with time during the rise of the Oct. 15 burst.
From the beginning of the burst, the amplitude decreases for $\sim
0.2$ s, and then assumes a near constant value, with some
fluctuations. This behavior is qualitatively similar to that seen in
bursts from the LMXB systems 4U 1728-34 and 4U 1636--536 (Strohmayer,
Zhang \& Swank 1997; Bhattacharyya \& Strohmayer 2005).  We also
perform time resolved spectral fitting (using a blackbody model)
during the rise of the Oct. 15 burst.  The inferred source radius
(which provides some relative indication of the size of the burning
region on the stellar surface) shows evidence for a modest increase
for the first $\sim 0.2$ s, and then remains almost constant (panel
{\it c}, Fig. 1). Panel {\it d} gives the corresponding source
temperature variation.

\section {Discussion} \label{sec: 3}

Taken together our results for the three bursts from SAX J1808.4--3658
suggest that the oscillation frequency can evolve in a complex manner
during burst rise.  In all cases a constant frequency model is a poor
description of the oscillations. In two bursts (Oct. 15 and 18) the
data can best be described by a complex modulation in frequency
whereby it initially increases, then decreases before increasing
again. Moreover, the oscillation amplitude and the inferred burning
region area are found to be correlated with the frequency.  We now
discuss from a theoretical perspective how spreading of thermonuclear
flames can plausibly account for these observations. In our
discussions we use the observational results mostly from the Oct. 15
burst as characteristic. The salient features of this burst are as
follows: (1) from the start of the burst, the oscillation frequency
and burning region area increase and the oscillation amplitude
decreases for $\sim 0.2$ s; (2) after $\sim 0.2$ s, the frequency
first decreases and then increases, and both amplitude and burning
region area reach a nearly constant value (with some fluctuations).

The burst begins when the fuel (i.e., accumulated matter) ignites at a
particular point, and then the flame propagates over the surface
(Fryxell \& Woosley 1982; Spitkovsky et al. 2002; Bhattacharyya \&
Strohmayer 2006a; 2006b).  Before spreading has engulfed the entire
star, temperature variations due to surface waves may not be able to
explain the brightness oscillation or its frequency evolution (as has
been proposed for the burst tail oscillations; Heyl 2005; Lee \&
Strohmayer 2005), as the rapid spreading and temperature increase may
wash out this effect. This explanation of oscillations during burst
rise was also shown to be unfavored by Bhattacharyya \& Strohmayer
(2005).  However, thermonuclear burning in a limited portion (hot
spot) of the neutron star surface can give rise to these oscillations,
and the propagation of the burning front may explain the observed time
evolution of oscillation properties.

In order to understand how thermonuclear flame spreading can give rise
to the observed evolution of oscillation amplitude and burning region
area, we first review some relevant results from previous work
(Spitkovsky et al. 2002): (1) the greater scale height of the burning
region than the cold fuel gives rise to a shearing speed (as the
horizontal pressure gradient in the burning front increases with
height). As a result, the cold fuel is drawn into the burning front
and ignited. This enables the burning front to propagate. (2) The
shearing speed is greater nearer the equator than the pole (due to the
latitude dependence of the Coriolis parameter; Spitkovsky et
al. 2002). Thus, the burning front propagates with the shearing speed
$(\vartheta)$ (assuming the mixing time scale is very small;
Spitkovsky et al. 2002; Fujimoto 1988; Cumming \& Bildsten 2000).
Now, if the fuel ignites at a mid-latitude (say, in the northern
hemisphere), it will propagate faster towards the equator (than the
pole), and the east-west width of the burning region will increase
much faster near the equator. Therefore, after a certain time, the
burning region will encircle the equator and propagate more or less
symmetrically towards the south pole. The northern burning front
propagates towards the north pole, keeping the asymmetry (due to the
variation of east-west width with latitude; see Figure 8 of Spitkovsky
et al. 2002), which vanishes near the pole. At the beginning, the
burning region is relatively small, and hence the oscillation
amplitude can be large.  As the burning region grows, the oscillation
amplitude naturally diminishes (Fig. 1). This effect was also reported
for bursts from other sources (Strohmayer, Zhang, \& Swank
1997). After the burning region encircles the equator with a
considerable north-south width, the observed burning area does not
increase much (hence the near constant radius after the initial
increase; see Fig. 1). From this time, the oscillation is due to the
residual asymmetry of the northern burning front (with the persistent
background due to the azimuthally symmetric portion of the burning
region), and hence the amplitude attains a near constant value till
the asymmetry vanishes. Therefore, according to our explanation, for
the Oct. 15 burst, the burning region takes $\sim 0.2$ s to nearly
encircle the stellar equator. However, we note that this explanation
does not include the effects of magnetic field, which may affect the
burning front propagation considerably (see below).

Thermonuclear flame propagation may give rise to the observed
frequency evolution, if the eastbound and the westbound burning fronts
have different accelerations, causing acceleration (either eastwards
or westwards) of the center of the burning region relative to the
star.  
Now considering this picture of the azimuthal shift of the hot spot
center, an observed oscillation frequency lower than the stellar spin
frequency may be caused by the westward motion of the burning region
center (for an eastward rotating star).  An increased scale height in
this region may cause such motion, because, as the hot portion of the
burning region puffs up, its top portion slips westwards to conserve
angular momentum. If the shearing speed due to this is $v$, the center
of the burning region moves westward with a speed $v$ relative to the
stellar surface. However, this effect alone can not explain an
oscillation frequency that is more than $\sim 2$ Hz lower than the
stellar spin frequency (Cumming \& Bildsten 2000). Therefore, as the
initial oscillation frequency of the Oct. 15 burst is $\sim 3-4$ Hz
less than the neutron star spin frequency, this physical effect can
only partially explain the initial frequency of this burst. Moreover,
the increased scale height effect can not explain the following
observed features: (1) initial quick increase of frequency (as the
scale height does not decrease during the beginning of burst rise),
(2) oscillation frequency overshooting the stellar spin frequency, and
(3) a decrease and subsequent increase of oscillation frequency during
the later stage of the burst rise. A physical effect of variable
magnitude, that can move the center of the expanding burning region
(hot spot) both eastwards and westwards, is required to explain these
aspects. This effect can add to the increased scale height effect at
the burst onset (if moving the hot spot center westwards), and can
compete with the scale height effect to produce the other observed
features of frequency evolution (if moving the hot spot center
eastwards). At present, such an effect is not known, but the surface
magnetic field may be a promising candidate. This is because the
surface magnetic field may have its strength amplified and its
geometry modified by differential rotation during the flame spreading.
This may enhance the magnetic force, which can subsequently modify the
shearing flows, and hence can influence the propagation of the burning
front (e.g., Cumming et al. 2002). The exact nature and magnitude of
this influence will depend on the geometry and the strength of the
field. This effect may be particularly important for SAX
J1808.4--3658, as this source is a pulsar (and hence probably has a
higher magnetic field than non-pulsars), and because the oscillation
frequency overshooting the neutron star spin frequency has been
observed only from this source. This source also seems different from
other sources in other properties such as the quick and large increase
of oscillation frequency during burst rise. However, we note that the
uniqueness of SAX J1808.4--3658 in terms of burst rise oscillation
frequency evolution is not yet well proven, as so far only seven
bursts with significant frequency evolution during the rising phase
have been observed (three from this source, and four from 4U
1636--536; Bhattacharyya \& Strohmayer 2005).  Nevertheless, detailed
numerical simulations of thermonuclear flame spreading (including
magnetic field effects) may be able to explain the observed correlated
evolution of oscillation properties during burst rise, and may provide
important information about how the flame speed and other flame
spreading properties can be influenced by stellar spin, compactness,
magnetic field and burst strength.  Therefore, we emphasize that burst
rise oscillations provide a potentially powerful tool to understand
thermonuclear flame spreading on neutron star surfaces under extreme
physical conditions.

\acknowledgments

{}

\clearpage

\begin{deluxetable}{cccccccc}
\tablecolumns{8} \tablewidth{0pc} \tablecaption{Frequency evolution
model parameters\tablenotemark{a} (with 1$\sigma$ error) for burst
rise oscillations of three bursts from SAX J1808.4--3658.}
\tablehead{Burst date & $\nu_{\rm 0}$ & $\dot\nu_{\rm 1}$ & $t_{\rm
b_{\rm 1}}$ & $\dot\nu_{\rm 2}$ & $\ddot\nu_{\rm 2}$ & $\chi^{\rm
2}$/dof & $Z^2_1$\tablenotemark{b}} \startdata 2002 Oct 15 &
$396.72^{+2.00}_{-21.73}$ & $29.64^{+236.21}_{-13.95}$ &
$0.20^{+0.05}_{-0.10}$ & $-15.54^{+4.73}_{-4.94}$ &
$23.67^{+11.52}_{-9.81}$ & $3.36/4$ & $152.16$ \\ \\ 2002 Oct 18 &
$398.22^{+0.20}_{-0.21}$ & -- & $0.31^{+0.04}_{-0.04}$ &
$19.21^{+5.19}_{-3.39}$ & -- & $8.44/6$ & $62.83$ \\ \\ 2002 Oct 19 &
$399.64^{+0.23}_{-0.14}$ & $1.69^{+0.07}_{-0.42}$ & -- & -- & -- &
$6.46/7$ & $86.64$ \\ \enddata \tablenotetext{a}{Frequency evolution
model: $\nu(t) = \nu_{\rm 0} + \dot\nu_{\rm 1} t$ (for $t \le t_{\rm
b_{\rm 1}}$, and $\nu_{\rm 0} = \nu(0))$; $\nu(t) = \nu(t_{\rm b_{\rm
1}}) + \dot\nu_{\rm 2} (t-t_{\rm b_{\rm 1}}) + \ddot\nu_{\rm 2}
(t-t_{\rm b_{\rm 1}})^2$ (for $t \ge t_{\rm b_{\rm 1}})$.}
\tablenotetext{b}{Fundamental power during burst rise.}
\end{deluxetable}

\clearpage
\begin{figure}
\hspace{-1.1 cm}
\epsscale{0.9}
\plotone{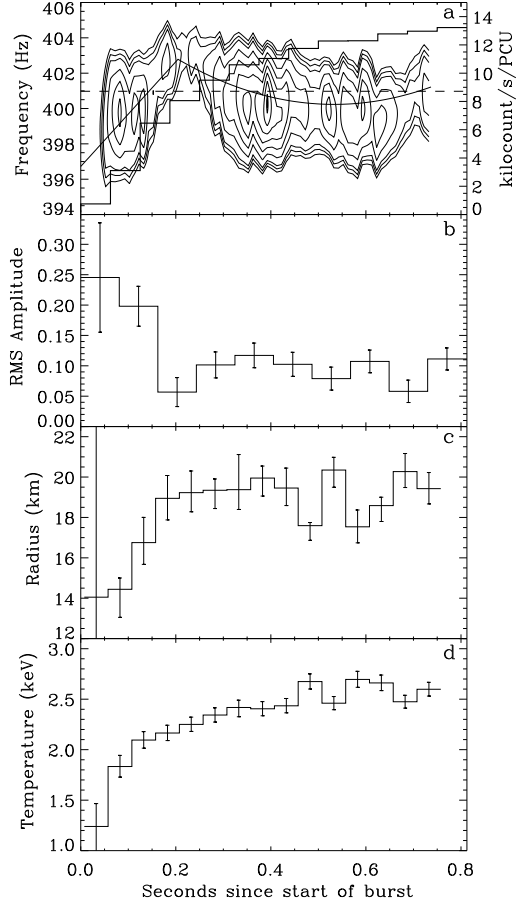}
\vspace{-3.5 cm}
\caption {Time evolution of different observed burst properties during
the rise of the Oct 15 burst from SAX J1808.4--3658. Panel {\it a}
gives the detected intensity (histogram), power contours (minimum and
maximum power values are $16$ and $51$) from the dynamic power spectra
(for 0.15 s duration at 0.01 s intervals), the best fit model from
Table 1, and the neutron star spin frequency (broken horizontal line).
Panel {\it b} shows the rms amplitude of the oscillations. Here the
horizontal lines give the binsize. Panel {\it c} gives the inferred
radius (assuming 10 kpc source distance) of the source, while panel
{\it d} gives the corresponding temperature. For panels {\it b}, {\it
c}, and {\it d}, persistent emission is subtracted and a deadtime
correction is applied, and the error bars are $1\sigma$ values.}
\end{figure}

\clearpage
\begin{figure}
\epsscale{0.8}
\plotone{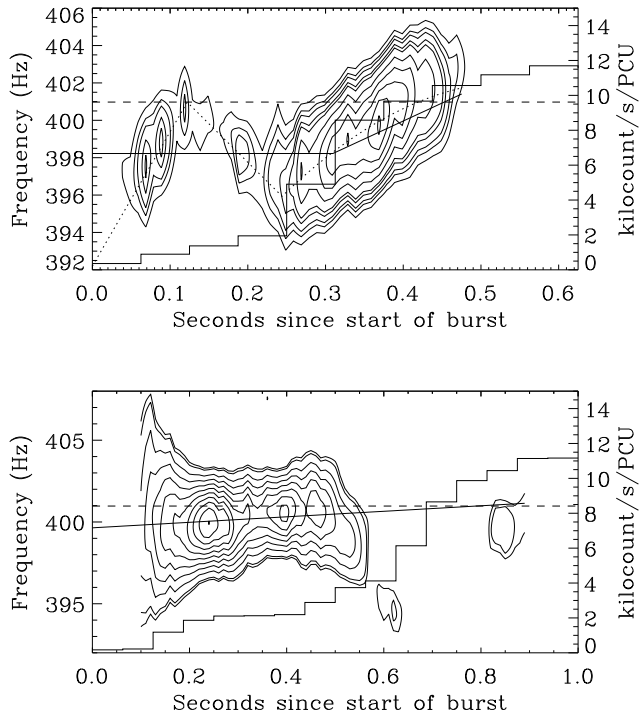}
\vspace{-3.0 cm}
\caption {Similar to panel {\it a} of Fig. 1 but for the Oct 18
(upper) and Oct 19 (lower) bursts. For the upper panel dynamic power
spectra are calculated for 0.15 s duration at 0.01 s intervals, while
for the lower panel these numbers are 0.2 s and 0.01 s. For the
calculation of power contours, minimum and maximum power values are
15 and 50 for the upper panel, and 17 and 111 for the lower panel. The
dotted line in the upper panel gives the frequency evolution for which
the total power $(74)$ is higher than that $(63)$ corresponding to the
best fit model (solid line) from Table 1.}
\end{figure}

\end{document}